\documentclass[review]{elsarticle}

\usepackage{lineno,hyperref}
\usepackage{array}
\usepackage{appendix}
\usepackage{hyperref}
\usepackage{url}
\hypersetup{pdfauthor={Name}}
\journal{Remote Sensing}

\biboptions{authoryear}

\begin{document}

\begin{frontmatter}

\title{Effects of the COVID-19 lockdown on urban light emissions: ground and satellite comparison}

\author[1]{M\'aximo Bustamante-Calabria\corref{mycorrespondingauthor}}\ead{maximo@iaa.es}
\author[1,2,3]{Alejandro S\'anchez de Miguel}
\author[1]{Susana Mart\'in-Ruiz}
\author[1]{J. L. Ortiz}
\author[1]{J. M. V\'ilchez}
\author[1]{Alicia Pelegrina}
\author[1]{Antonio Garc\'ia}
\author[3]{Jaime Zamorano}
\author[4]{Jonathan Bennie}
\author[2] {Kevin J. Gaston}
\address[1]{Instituto de Astrof\'isica de Andaluc\'ia, Glorieta de la Astronom\'ia, s/n, C.P.18008 Granada, Spain}
\address[2]{Environment and Sustainability Institute, University of Exeter, Penryn, Cornwall TR10 9FE, U.K.}
\address[3]{Depto. F\'isica de la Tierra y Astrof\'isica. Instituto de Física de Part\'iculas y del COSMOS (IPARCOS), Universidad Complutense, Madrid, Spain}
\address[4]{Centre for Geography and Environmental Science, University of Exeter, Penryn, Cornwall TR10 9FE, U.K.}
\begin{abstract}
'Lockdown' periods in response to COVID-19 have provided a unique opportunity to study the impacts of economic activity on environmental pollution (e.g. NO$_2$, aerosols, noise, light). The effects on NO$_2$ and aerosols have been very noticeable and readily demonstrated, but that on light pollution has proven challenging to determine. The main reason for this difficulty is that the primary source of nighttime satellite imagery of the earth is the SNPP-VIIRS/DNB instrument, which acquires data late at night after most human nocturnal activity has already occurred and much associated lighting has been turned off. Here, to analyze the effect of lockdown on urban light emissions, we use ground and satellite data for Granada, Spain, during the COVID-19 induced confinement of the city's population from March 14 until May 31, 2020. We find a clear decrease in light pollution due both to a decrease in light emissions from the city and to a decrease in anthropogenic aerosol content in the atmosphere which resulted in less light being scattered. A clear correlation between the abundance of PM10 particles and sky brightness is observed, such that the more polluted the atmosphere the brighter the urban night sky. An empirical expression is determined that relates PM10 particle abundance and sky brightness at three different wavelength bands.
\end{abstract}


\end{frontmatter}

%\linenumbers

\section{Introduction}
The COVID-19 pandemic has caused dramatic changes in human habits and activities across much of the world. This has been especially true during so-called 'lockdown' periods, when the local, regional or national activities and movements of people have been markedly curtailed to reduce rates and levels of viral transmission. The details of these restrictions, and the extent and vigour with which they have been enforced, have varied greatly between different countries. Nonetheless, these unusual situations have acted as valuable 'natural experiments', allowing novel analyses to be conducted of the relationships between levels of human activity and levels of damage to the environment, by comparing important potential impacts prior to and during lockdown periods. For example, this has been done for NO$_2$ concentrations in urban areas and globally for aerosol content \citep{nasaearth, nasaair}, noise \citep{nyt}, and seismic tremors \citep{xiao2020covid19}. Also, the impact of lockdown on night sky brightness in Berlin has been studied \citep{Jechow2020Berlin}.

Many of the environmental impacts of human activity, and hence the effects of the lockdown periods, can be monitored remotely through satellites. However, perhaps surprisingly, this has proven challenging to do at large regional or global scales for nighttime light emissions, which have been recognised as both a valuable indicator of human population density, urbanisation and economic activity \citep{small2013night,levin2017global}, and also as themselves having important impacts on human health and the natural environment \citep{rich2013ecological,gaston2013ecological,koo2016outdoor,falchi2011limiting}. The primary source of remote sensing data on nighttime light emissions is the Suomi-North Polar Partnership/Visible Infrared Imaging Radiometer Suite-DayNightBand (SNPP/VIIRS-DNB). This provides data of intermediate spatial resolution using a panchromatic sensor, but its observation time is after 01:30 local time limiting its usefulness for determining changes in nighttime lighting at times when people are predominantly active. Sensors are also carried on other satellite platforms, but variously have limited spatial coverage, require substantial data calibration, and/or are private and with data costs that are prohibitive for monitoring for research purposes \citep{levin2020remote}. Images of the earth taken by astronauts aboard the International Space Station (ISS) using DSLR cameras are also available, have high resolution and are multi-spectral, but are not systematically acquired in space or time, also greatly limiting their suitability.

Accepting these limitations on a broad scale places disproportionate significance on ground-based measurements of changes in artificial nighttime lighting as a means of determining the effects of COVID-19 lockdown periods. Unfortunately, the availability of such data is quite limited, although there are ongoing attempts both to build and maintain networks of monitoring sensors \citep{cheung2015globe,zamorano2016stars4all} and to provide platforms to encourage and collate regular spot measurements by citizen scientists \citep{walker2008globe}. 

Here we determine the impacts of a COVID-19 lockdown on artificial nighttime light emissions for an exemplar, the city of Granada, Spain, using both satellite and ground-based measurements.  In Spain the general lockdown, which started on March 14th 2020, was particularly severe to counteract a steep rate of spread of the infection. 

\section{Methods}
\subsection{Satellite data}
For satellite images of Granada we used the SNPP-VIIRS/DNB VNP46A1 product \citep{roman2018nasa}. Details on the image processing undertaken are provided in \cite{roman2018nasa}, but briefly, images were corrected for atmospheric, topographic and cloud effects. Radiometric calibration was carried out but seasonal and moon effects were not taken into account. Therefore, for this paper, images with no moon illumination were selected, considering only those taken two days prior or two days after the new moon, and only images obtained in January to May were used. Only 27 images in total fulfilled these selection criteria: 5 images for 2018, 7 images for 2019 and 14 images for 2020. Of this last group, 6 images were obtained before the lockdown and 8 during the lockdown.  In some months, such as April, only two images were available, due to cloud cover \citep{AEMET}. The images were manually inspected to detect any kind of cloud or fog features (blurriness) or anomalous dimming that could be explained by image acquisition at very shallow angles. The photometry was performed using "FunTools" \citep{mandel2011funtools} and the visualization was done with SAO-DS9 \citep{joye2003new}. 

\subsection{Ground-based sky brightness measurements}
In order to measure sky brightness from the ground we used a set of Sky Quality Meters (SQMs; Unihedron). SQMs (\cite{cinzano2005,hanel2018}) measure night sky brightness and have been used in a large number of studies directly or indirectly concerned with light pollution (\cite{falchi2016new,kyba2015,doi:10.1093/mnras/stx145}). They have a spectral response from 320–700 nm, approximately overlapping the Johnson B and V bands used in astronomical photometry. We used the SQM-LE model, a lensed version with Ethernet connection whose lens reduces the angular sensitivity to $\sim$20º (full width at half maximum, FWHM) around the zenith. The SQM photometers directly provide data in $mag/arcsec^{2}$ with a systematic error of 10 per cent (0.1 $mag/arcsec^{2}$) according to the manufacturer. All these sensors were cross calibrated with the All-Sky Transmission MONitor (ASTMON) for their respective bands. As \cite{cinzano2005} have showed, the SQM with B Johnson and V Johnson filters are very good match to the pure B Johnson and V Johnson filters astronomical filters. So, this calibration error is mainly for the non filtered SQM and not for the cross calibrated ones, that probably can achieve better performance. Note that the $mag/arcsec^2$ scale used in astronomy is somewhat counter-intuitive for people outside the field because it is logarithmic, and because the brighter the sky the smaller the value. 

The set of SQMs consisted of three devices situated on the roof of the Instituto de Astrof\'isica de Andaluc\'ia-CSIC (IAA-CSIC) (lat = N 37.1616º, long = W 3.59036º, h = 685 meters above mean sea level) located inside the city of Granada. Two were equipped with specific B and V Johnson band filters, whereas the other one was used without any filter. All devices were weatherproof to protect them from outdoor conditions. Each SQM sent data through the internet to a Raspberry Pi unit for storage and time-tagging and made about 1000 recordings of sky brightness over a night. We used the open-source software PySQM \citep{nievas2014} for these operations.

\begin{figure}[ht]
\begin{center}
\includegraphics[scale=0.5]{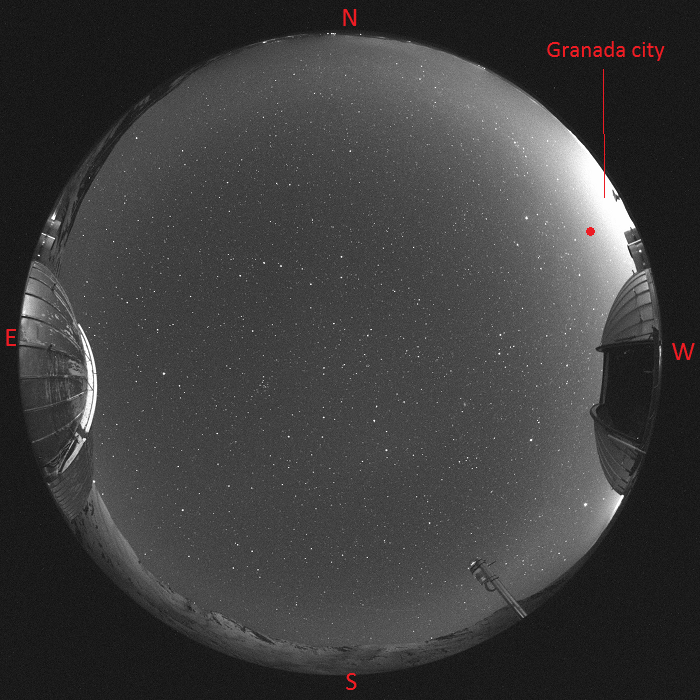}
\caption{Image from the ASTMON FoV in the V filter on February 20th 2020 at 01:16 UT. The red circle highlights the region of the celestial sphere where sky brightness measurements were taken. The instrument is situated on the roof of the main building in the middle of the two domes.}
\label{fig:astmon}
\end{center}
\end{figure}

In addition, the Sierra Nevada Observatory, located more than 2000 meters above and more than 20 km away from the city of Granada (lat = N 37.0642º, long = W 3.38472º, h = 2900 mamsl), is equipped with an ASTMON (All-Sky Transmission MONitor), an imaging device that measures sky brightness in the UBVRI Johnson bands at all-sky locations \citep{aceituno2011}, not just at the zenith. The instrument is fully robotic, i.e., it performs all astronomical data reduction and calibration tasks automatically, obtaining maps of sky brightness through the night. For this study, we selected the values of sky brightness at an azimuth and altitude where the glow of Granada's city is clearly observed ($\sim$25º above the horizon) as shown in figure \ref{fig:astmon}. Locations are shown in figure 2.

\begin{figure}[ht]
\begin{center}
\includegraphics[scale=0.4]{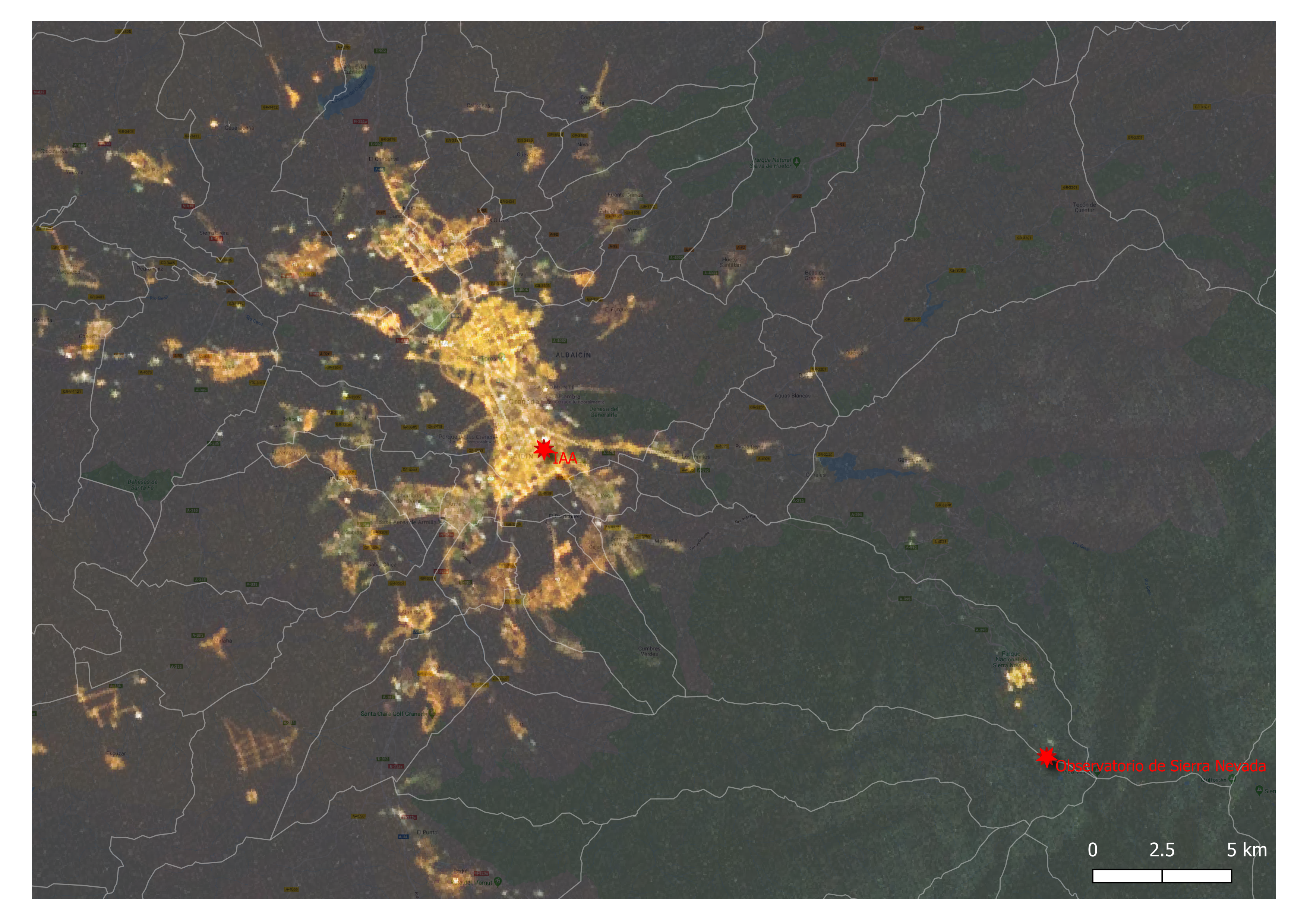}
\caption{Image of Granada city and metropolitan area taken from the International Space Station on February 6th 2012 and geo-referenced, with the gray lines delimiting different administrative regions. The red stars indicate the positions of the sensors used in this paper.}
\end{center}
\label{fig:fig2}
\end{figure}

We analyzed data from January, February and until March 14 2020 to represent the conditions prior to the lockdown, and data from March 14 until May 23 to represent the lockdown conditions. Sky brightness data are extremely sensitive to cloud cover and lunar phase. In astronomy, sky brightness measurements are reported only for moonless nights and for completely cloudless nights. For this study, we did the same. Cloudless nights were selected following the criterion that there was no flickering in the measurements with a standard deviation of 0.10 mag. Sky brightness is usually measured at the zenith. Unfortunately, not all of the measurements reported in the literature specify whether these are of the sky background (removing the stellar contribution) or of the night sky containing stars. SQMs do not correct for starlight contribution. The stellar contribution might seem negligible but it is not, at least for regions with little light pollution, and this is especially relevant when the Milky Way or other dense stellar regions fall within the viewing cone of SQMs. To estimate this "seasonal" effect from our measurements, we analyzed SQM data from our remote observatory at La Sagra (lat = N 37.98º , long = W 2.56º, h = 1530 mamsl) which has minimal light pollution in order to determine the amount of change from the period of January-February to the lockdown period. We used data for 3 years to have a robust estimation that would not be affected by weather conditions. See \ref{A:1}.

\subsection{Pollution data}
Atmospheric pollution data were obtained from the measuring station of the Granada Congress Palace, published by the Junta de Andalucía and available on the Granada City Council website \citep{calidadaire}. These data were used as the station is the closest to the IAA headquarters (coord. UTM: X = 446721, Y = 4113421).  We focused on the concentration of PM10 particles (particles with diameter $< 10 \mu m$) and nitrogen dioxide, two pollutants directly related to vehicle emissions and therefore to urban activity. The concentration values ($ \mu g/m^{3} $) were obtained every hour, so in order to study possible correlations with the brightness of the sky we selected the measurements of the SQM devices installed in the IAA-CSIC and the ASTMON of the OSN (at a point approximately 25º above the city of Granada) with different filters, at valid nights and coinciding in time with the measurements of the concentration of particles and nitrogen dioxide. Thus we gathered a sample of 137 measurements in the night hours distributed from January to May 2020.

\section{Results}
\subsection{Satellite data}
\begin{figure}[ht]
\includegraphics[scale=0.2]{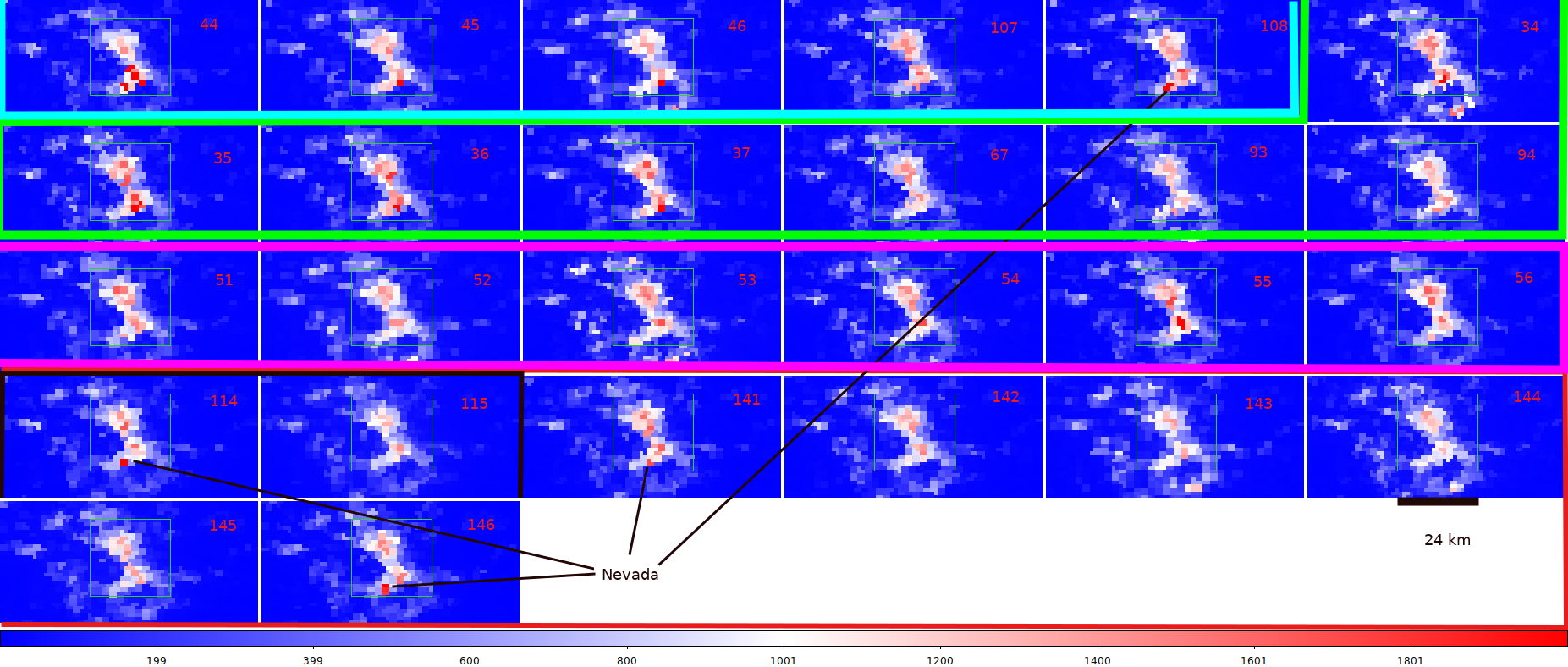} \caption{SuomiNPP/VIIRS-DNB observations of the Granada region using the VNP46A1 product. Cyan images are for 2018, green images for 2019. Magenta images are before the lockdown 2020 and red images for nights during the lockdown. The main isolated light pollution source in Granada, the Nevada commercial center, is marked, although it is not always visible from the satellite images because of different viewing angles on different nights.}
\label{fig:VIIRS}
\end{figure}

Figure \ref{fig:VIIRS} shows a comparison of satellite imagery of Granada on clear and moonless nights prior to lockdown and during lockdown. Quantitative analysis indicates that the average intensity of emissions decreased by around $10\% \pm 5\%$ during the lockdown. Comparison of 2018 and 2019 data indicate similar emissions values, although Granada has an ongoing municipal program of transition from High Pressure Sodium (HPS) to LED lighting, so this can only be considered as a lower threshold of change. The dispersion of the day to day intensity observations is of 14\%, although it is worth mentioning that the lowest values of the series occurred during the lockdown. The brightest individual light source in Granada (indicated as "Nevada" in the figure), which was lit up constantly during the lockdown is not visible on many of the images because of the angle of observation. This illustrates the important role of the angle of observations in these measurements. Some other research has not consider this potential issue \citep{liu2020spatiotemporal,elvidge2020dimming,ghosh2020dimming}.

\subsection{Ground-based data}
\subsubsection{Measurements comparing two time intervals on different nights}\label{S:3.2.1}
As a first approach to analysing the SQM data,  values were compared for two time intervals, one before midnight (from 21:00 to 22:00 UT) and another one after midnight (from 2:00 to 3:00 UT). This has been done to see if the changes in lighting and those associated with the decrease in activity are reflected in any way in the values of sky brightness in different filters, taking into account that part of the ornamental lighting is turned off in the second half of the night and that the intensity of the general lighting decreases after legal midnight. Figure 4 shows the average values of sky brightness from the IAA headquarters in the two time intervals mentioned and at the valid nights, and the average of the B-V colour index which results from subtracting the measurements obtained with the two filters in these time intervals.

\begin{figure}[ht]
\begin{center}
\includegraphics[scale=0.255]{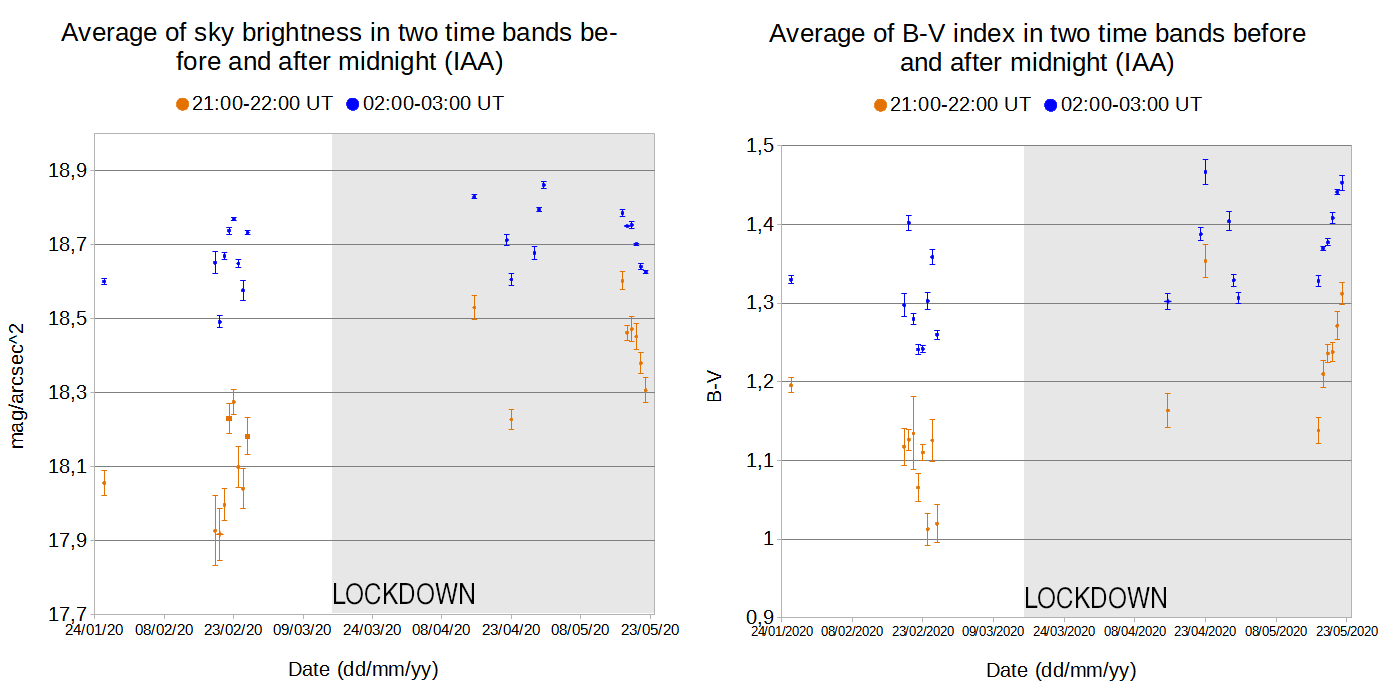}
\caption{Left: time-averaged sky brightness from the unfiltered SQM photometer as a function of date. Right: time-averaged B-V index. Photometers located at IAA headquarters in Granada. The pre-lockdown brightness is much higher than in the lockdown period, especially in the hour interval before midnight (brown plus symbols). Note that we are using the standard surface brightness units in astronomy, and the magnitude scale decreases for higher brightness.}
\end{center}
\label{fig:SQM1}
\end{figure}

We have defined two groups of data based on the pandemic's policies in Spain: from the months of January (26th) and last week of February (19th to 26th); and from the months of April (15th, 22nd, 23rd, 28th, 29th and 30th) and May (17th to 22nd). The first group corresponds to pre-lockdown sky brightness and the second to sky brightness obtained at the time of lockdown. The first thing to note is that both in the brightness measurement and in the B-V index there is a clear difference between the first hours of the night and the last hours, especially before lockdown: from 21:00 to 22:00 UT the night sky of Granada is on average 0.57 $mag/arcsec^{2}$ brighter than from 2:00 to 3:00 UT. Similarly, the average value of B-V is higher from 2:00 to 3:00 UT (B-V 1.30 before midnight and 1.10 after midnight before the lockdown, so a 0.2 mag difference).

The variation in the colour index suggests that this is a consequence of the turning off of ornamental lighting (examples are the Alhambra Palace illumination and facade illuminations of several monuments) and private lighting (aka. cars, private outdoor lighting, commercial lighting and indoor lighting), as most of the lamps used for ornamental lighting and private outdoor lighting are metal halide lamps, or have been replaced by LED that produce white or blue-white light with a significant emission in blue, so that once they are switched off the records of the photometer with B filter are significantly higher (due to the lower brightness in this band). This effect has been documented before in many other cities, like Berlin or Madrid (\cite{kyba2012red,de2015variacion,bara2019estimating}).

This difference between night hours also occurred during the lockdown, although to a lesser extent. During lockdown, the Granada sky between 21:00 and 22:00 UT was on average 0.31 $mag/arcsec^{2}$ brighter than 2:00 to 3:00 UT, while in the B-V index there was a difference of 0.14. 

If we compare the days before and during the lockdown, the Granada sky between 21:00 and 22:00 UT was 0.33 $mag/arcsec^{2}$ darker after its declaration; the difference was 0.08 $mag/arcsec^{2}$ from 3:00 to 4:00. For the B-V index the differences were greater in the first half of the night (0.14) than in the early morning hours (0.08). 

\subsubsection{Evolution of measurements during the night. Average nights before and during lockdown}
Instead of analysing the average values in time periods before and after midnight, the average variation between 23:00 and 4:00 (legal time: UTC+1) of the brightness of the Granada sky and its colour index can be compared between February and April and May (legal time UTC+2). The average curve for the last week of February represents a more or less progressive darkening over the course of the night from 18.2 $mag/arcsec^{2}$ to over 18.6 at 4:00, with a steeper slope until midnight and then smoothing out (see figure 5). There are three steps or ``jumps" in the  curve: the first (and greatest) occurs at about 23:00; the second occurs at 00:00 (midnight legal time) and is less noticeable than the other two; and the third one occurs at 2:00. If we look at the B-V curve, the 2:00 step appears while the others are not so clear. It also appears on the curve for the days following the lockdown (summer time: UTC+2). We can infer that at that time some important lighting with a considerable emission in the blue band is switched off . The other steps may be related to a decrease in the intensity of public lighting, although in B-V a slight darkening can also be seen at midnight. 

The curves for the days following the lockdown show a greater divergence from the previous period in the early hours of the night, with higher values both in darkness (between 18.4 and 18.6 $mag/arcsec^{2}$) and in B-V (close to 1.3). In this case the legal midnight step is much more pronounced. If this is due to a decrease in the intensity of street lighting, it is interesting that before the lockdown this was not so clearly seen. The behaviour of the B-V graph can give some clues: before the lockdown it starts from values between 1.1 and 1.2 without exceeding 1.2 until after midnight, while in April it stays close to 1.3 until it reaches 1.4 in the second half of the night. In the first case, there is a greater brightness in filter B, which also decreases progressively as the night progresses, producing only a jump of some importance at 2:00. 

\begin{figure}[ht]
\begin{center}
\includegraphics[scale=0.78]{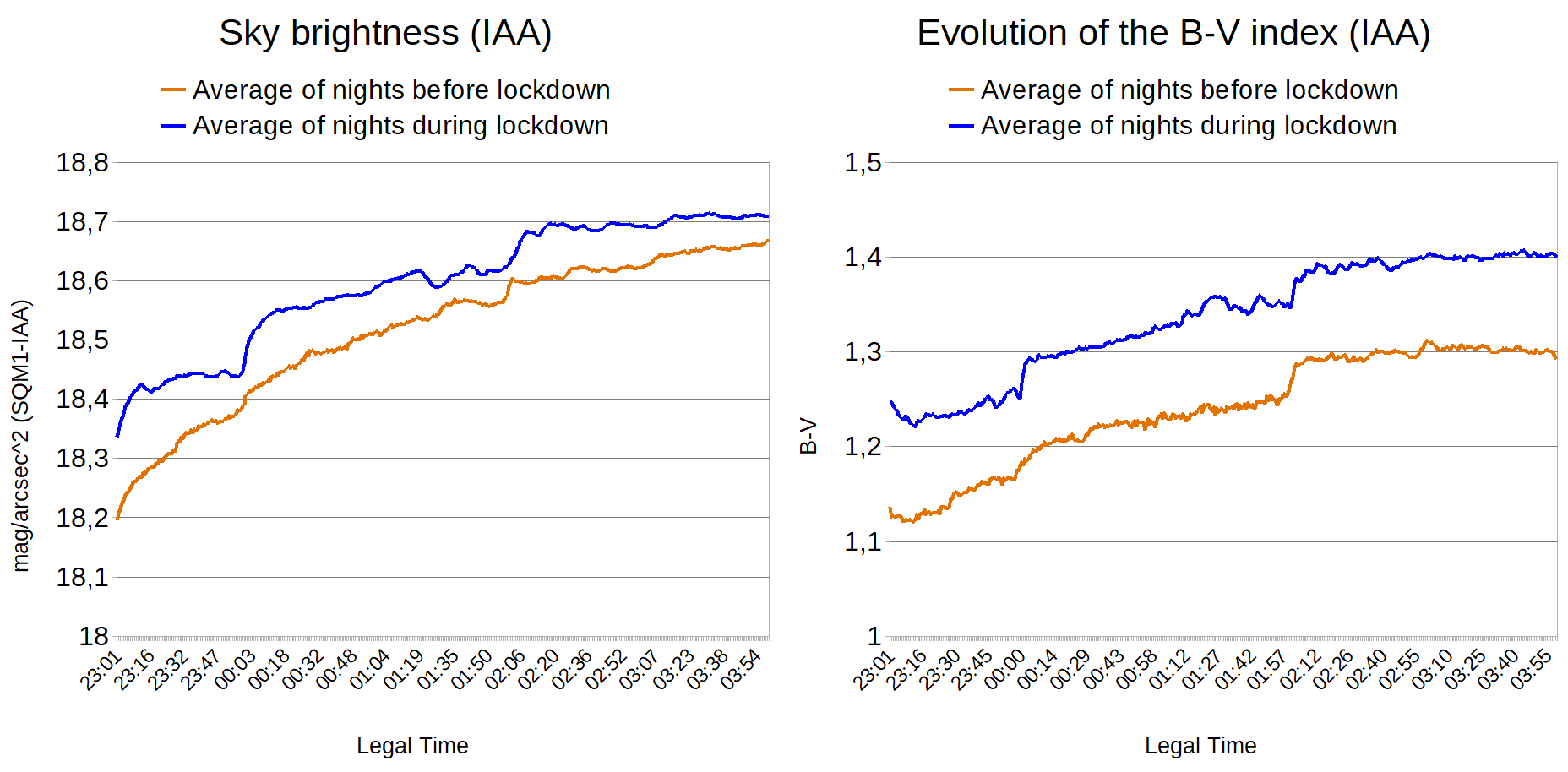}
\caption{Left: night-averaged sky brightness from the unfiltered SQM photometer as a function of time. Right: night-averaged B-V index. Photometers located at IAA-CSIC headquarters in Granada.}
\end{center}
\label{fig:SQM3}
\end{figure}

\subsubsection{Evolution of air pollution and sky brightness}
\begin{figure}[ht]
\begin{center}\includegraphics[scale=0.75]{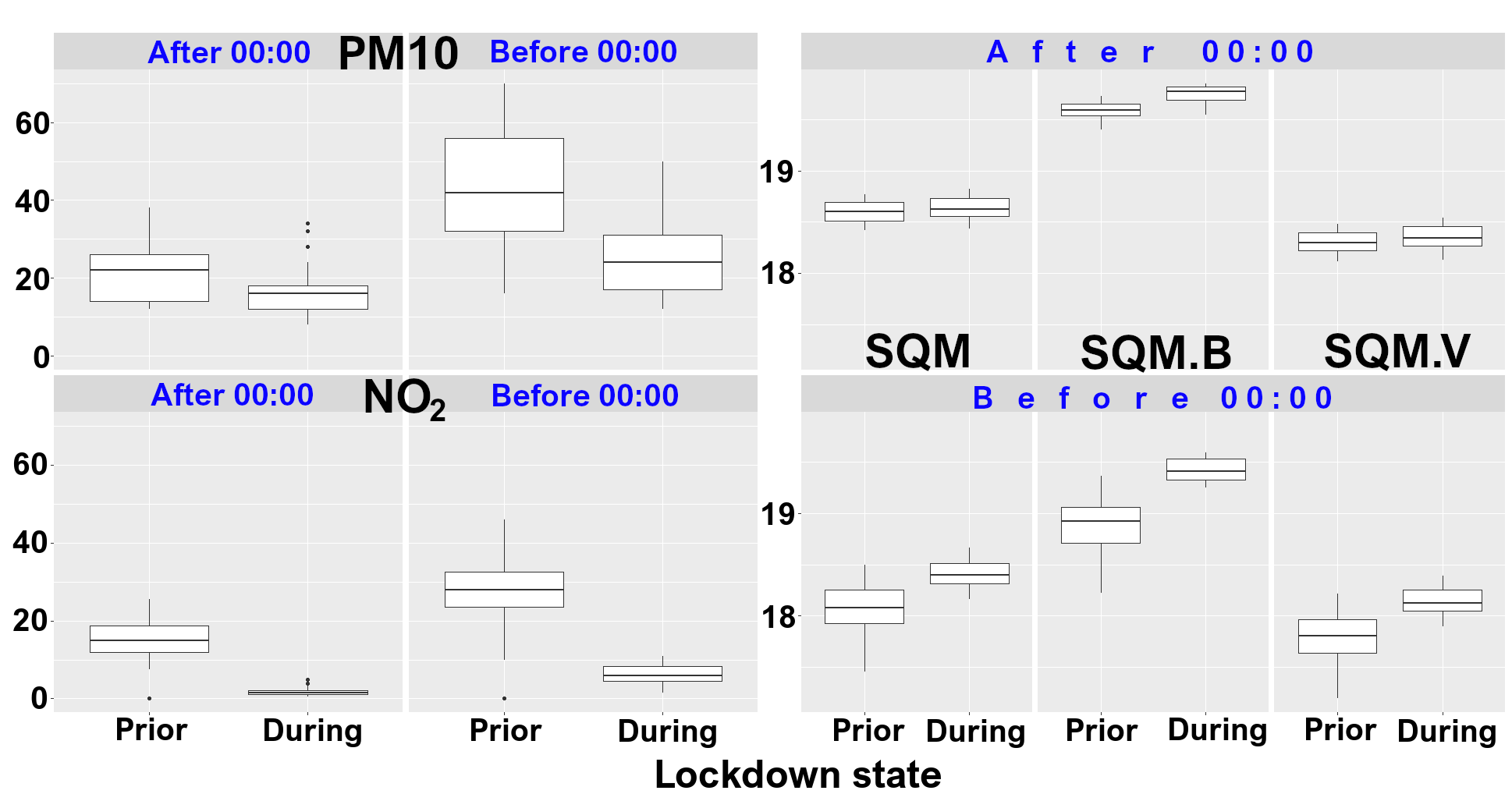}\end{center}
\caption{Left: particulate and nitrogen dioxide concentrations (in $ \mu g/m^{3} $) depending on time and lockdown state. Right: sky brightness (in $ mag/arcsec^{2} $) depending on time and lockdown state (IAA-CSIC, Granada).}
\label{fig:box}
\end{figure}

Both particulate PM10 and nitrogen dioxide concentrations were highest in the hours before midnight of the days prior to lockdown, while the lowest values occurred in the early morning and during lockdown (Figure \ref{fig:box} and table 1). Similarly, the Granada sky was darker at zenith during the early morning hours on days of lockdown and, conversely, brighter during the first nighttime hours prior to lockdown. The differences are less if we compare the hours after 00:00. In the case of measurements obtained with the ASTMON device the differences are less significant. Only in the B band a darkening of 0.12 $ mag/arcsec^2 $ is observed comparing the first hours of the night before and after the declaration of the lockdown.

\begin{table}\label{table:1}
\begin{center}
\caption{Average air pollution values and average sky brightness depending on time slot and lockdown state. (P.L. = Prior to lockdown; D.L. = During Lockdown; B.00:00 = Before legal midnight; A.00:00 = After legal midnight; N.F. = No Filter)\newline}
\begin{tabular}{ ||m{1.5cm}| |m{0.8cm} m{0.8cm}| m{0.8cm} m{0.8cm} m{0.8cm}| m{0.8cm} m{0.8cm} m{0.8cm}|| }
\hline
\hline
  & \multicolumn{2}{m{1.6cm}|}{\small $ \mu g/m^{3} $} & \multicolumn{3}{m{3cm}|}{\small SQM IAA ($ mag/arcsec^{2} $)} & \multicolumn{3}{m{3cm}||}{\small ASTMON OSN ($ mag/arcsec^{2} $)} \\
  & {\small PM10} & {\small $ NO_{2} $} & {\small N.F.} & {\small B} & {\small V} & {\small B} & {\small V} & {\small R} \\
\hline
\hline
{\small P.L./ B.00:00} & 43.32 & 27.59 & 18.06 & 18.89 & 17.78 & 20.71 & 19.56 & 18.91 \\
{\small P.L./ A.00:00} & 21.50 & 14.11 & 18.60 & 19.58 & 18.30 & 21.07 & 19.79 & 19.08 \\
\hline
{\small D.L./ B.00:00} & 25.37 & 6.13 & 18.40 & 19.42 & 18.14 & 20.83 & 19.54 & 18.86 \\
{\small D.L./ A.00:00} & 16.98 & 1.87 & 18.64 & 19.75 & 18.35 & 21.08 & 19.86 & 19.19 \\
\hline
\hline
\end{tabular}
\end{center}
\end{table}

The strongest correlations occur between the concentration of PM10 particles and the brightness of the sky SQM without filter and SQM with filter V ($ \rho = -0.84 $), where $\rho$ is the Spearman correlation index (see tables B.2 and B.3 in \ref{A:2}). Also noteworthy is the correlation between nitrogen dioxide concentration and sky brightness in the B-band (\textit{SQMB}) ($ \rho = -0.81 $), and with the B-V colour index ($ \rho = -0.77 $). The variables most related to the hour of the night (aka. proxy of human activity) of measurement are those corresponding to the sky brightness in all filters. This effect would dominate (higher correlation) versus air pollution (particle concentration).

In the case of ASTMON device measurements, there are also correlations between sky brightness and air pollution variables, although they are weaker than those described above. The best correlations occur between the measurements obtained in the B band and the concentrations of NO$_{2}$ ($ \rho = -0.66 $) and PM10 ($ \rho = -0.64 $). Although these results are significant, we have focused on the measures obtained within the city of Granada, as they present stronger correlations, especially for PM10 particle concentration. 

Figure 7 presents the measurements of PM10 particle concentration versus the sky brightness value (SQM without filter). The upper left hand area of the graph (darker sky and lower particle concentration) is mostly occupied by measurements taken after 00:00 (legal time) during lockdown, to which the lower NO$_{2}$ concentration values correspond. In contrast, the lower right zone (higher particle concentration and brighter sky) corresponds to hours before midnight on days prior to the declaration of the alarm state, and which are associated with higher NO$_{2}$ concentration values.

The linear fitting equations for measurements of sky brightness within Granada on different filters and the PM10 particle concentration values are: 

\begin{center}
\begin{tabular}{ m{4cm} m{6cm} }
SQM (no filter) & $ f(x) = 18.91(2) - 0.0179(8)x $ \\
SQM (filter V) & $ f(x) = 18.62(2) - 0.0176(8)x $ \\
SQM (filter B) & $ f(x) = 20.02(4) - 0.0227(1)x $ \\
\end{tabular}
\end{center}
\begin{center}
\small {(f(x): sky brightness in $mag/arcsec^2$; x: PM10 particle concentration in $\mu g/m^3$})
\end{center}
See tables C.4, C.5 y C.6 (\ref{A:3}) for errors, residuals and F-statistic. 

\begin{figure}[ht]
\begin{center}\includegraphics[scale=1.8]{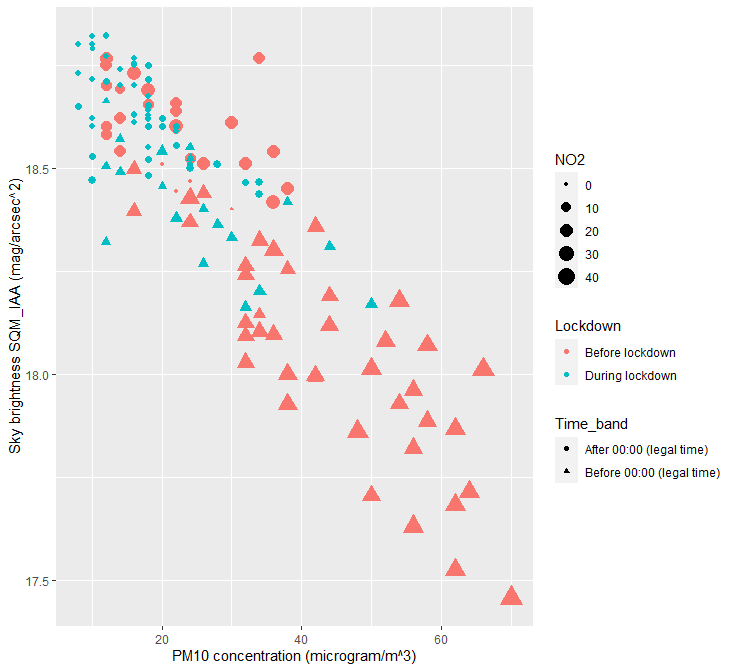}
\caption{Sky brightness within Granada (in $ mag/arcsec^{2} $) versus PM10 particulate concentrations (in $ \mu g/m^{3} $)  for the period prior to lockdown (pink symbols) and during lockdown (blue symbols) depending on time slot (triangles for the first half of the nights, dots for the second half of the nights). The PM10 concentration and SQM values are coinciding in time (every hour).}
\end{center}
\label{fig:corr1}
\end{figure}

The correlation between sky brightness and particulate air pollution is also evident by performing a multivariate analysis, including time as a third dimension. Variations in urban lighting depend mainly on the time of night, and therefore can be expected to influence the sky brightness values. On the other hand, pollution levels depend on human activity, which varies throughout the night. Thus, we have calculated a model that estimates a value of sky brightness as a function of time (as a fraction of a Julian day) and particle concentration. Tables C.7, C.8, C.9 and C.10 (\ref{A:3}) show the values of the multivariate models for sky brightness with SQM without filter and SQM with B filter, before and during lockdown. Figure 8 shows the model fitting for the unfiltered SQM photometer. 

\begin{figure}[ht]
\begin{center}\includegraphics[scale=0.9]{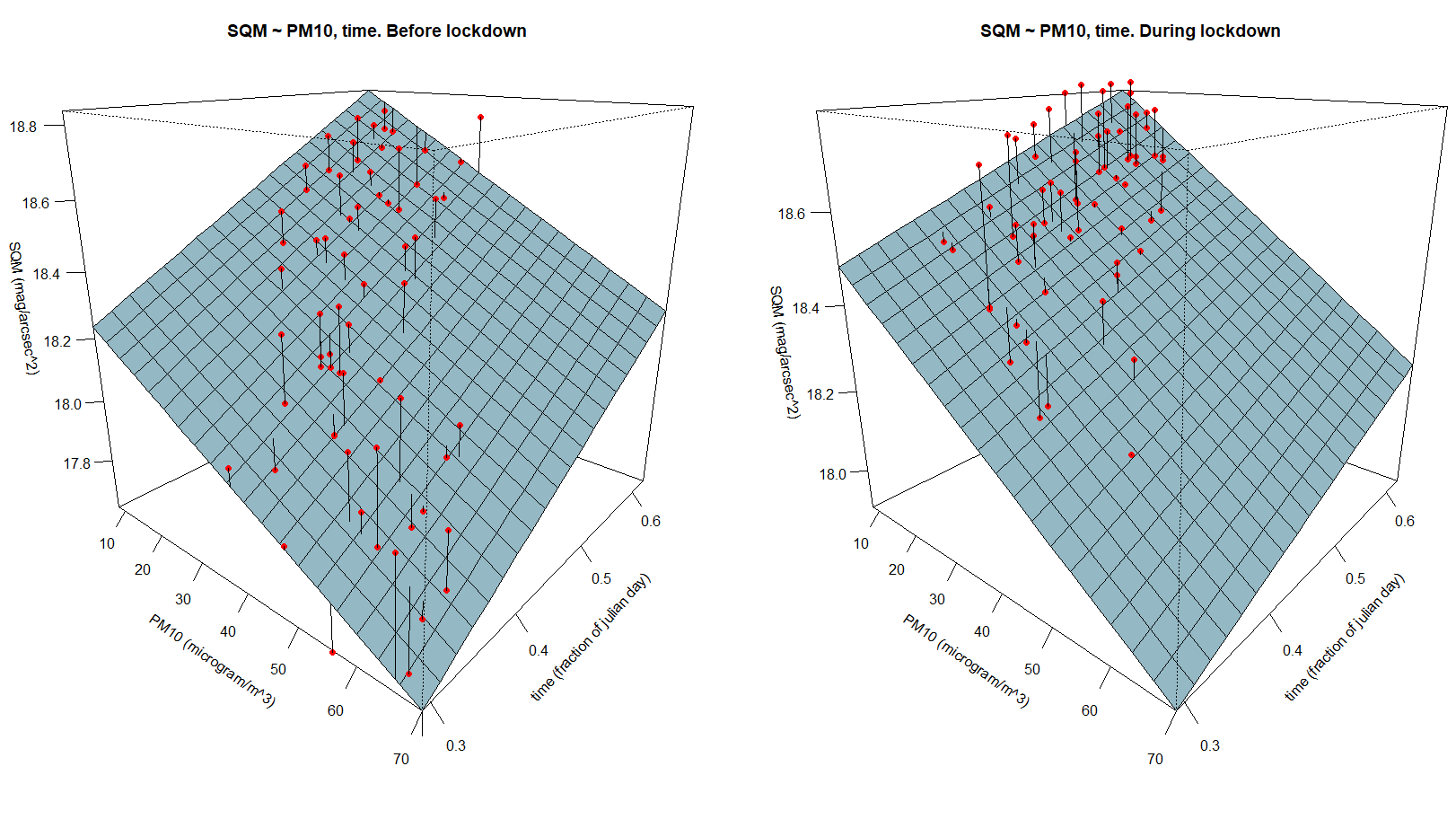}
\caption{Graphic representation of the multivariate lineal models between sky brightness (SQM), PM10 particle concentration and time (left: prior to lockdown; right: during lockdown)}
\end{center}
\label{fig:model1}
\end{figure}

The multivariate linear fitting equations for measurements of sky brightness within Granada on different filters, PM10 particle concentration values and time are:

\begin{center}
\begin{tabular}{ m{5cm} m{6cm} }
\small {SQM (no filter), prior to lockdown:} & $ f(x) = 17.79(4) - 0.0097(4)x + 1.79(2)t $ \\
\small {SQM (no filter), during lockdown:} & $ f(x) = 18.29(9) - 0.009(1)x + 0.94(4)t $ \\
\small {SQM (filter B), prior to lockdown:} & $ f(x) = 18.19(6) - 0.008(2)x + 2.76(5)t $ \\
\small {SQM (filter B), during lockdown:} & $ f(x) = 18.87(8) - 0.003(1)x + 1.71(3)t $ \\
\end{tabular}
\end{center}
\begin{center}
\small {(f(x): sky brightness in $mag/arcsec^2$; x: PM10 particle concentration in $\mu g/m^3$; t: time as a fraction of a julian day. See tables C.7, C.8, C.9 y C10 (\ref{A:3}) for errors, residuals and F-statistic)}
\end{center}

\section{Discussion}

\subsection{Correlation of sky brightness and PM10 particle abundance}

In figure 7 we show the sky brightness as a function of PM10 particle abundance measured in Granada from a station located around 1 km away from the site where the sky brightness measurements are taken. The altitude above sea level of the site is almost identical to that of the light pollution station.

Note that in that plot, the sky brightness measurements have not been corrected by the ``seasonal" effects of the milky way entering the field of view of the SQMs, but as pointed out in \ref{A:1} this can reach a 6\% contribution at most.

The correlation is obvious in the plot. Although the sky brightness can depend on many factors, it is clear that one of the main factors is the aerosol content. This is something that was expected from a physical point of view.  

\cite{garstang1991modeling} explicitly mentioned that the sky brightness seen close to the center of a city increases with the atmospheric aerosol content, according to \cite{garstang1986model} simple radiative transfer models. Indeed, this is what we observe in figure 8. The correlation of the sky brightness in magnitudes is linear with the PM10 concentration, which also makes sense given that the single scattering flux is proportional to $e^{\tau}$, where $\tau$ is optical thickness and $\tau$ is proportional to the particle concentration.

\subsection{Disentangling the effect of the net reduction in light emission from the effect of the decreased aerosol content}

We have seen that sky brightness depends both on aerosol content and on the time since the beginning of the night (the higher the aerosol particle concentration the higher the brightness and the earlier at night the higher the brightness). 
In this section we analyze the two correlations in terms of linear correlation in two variables and we have done that with data prior to lockdown and during lockdown.

From those fits, we can determine the night sky brightness for a reference particle concentration of 30 $\mu g m^{-3}$ and at 19.2 UT ($t = 0.3$ as a fraction of a julian day), prior to lockdown and during lockdown. The values are  18.04 mag/arcsec$^2$ and 18.29 mag/arcsec$^2$ before and during the confinement respectively. This means that at the same aerosol content and same time at night conditions, the confinement period was darker by 0.25 mag /arcsec$^2$ due to a decrease of the light output from the city. If we take into account that the brightness of the sky is around 5$\%$ higher in the period of time prior to lockdown due to higher star density getting in the detector, an approximate estimation of the decrease in the light output of the city is around 20 per cent. This percentage is probably related to the reduction in light from vehicles and from private lighting.

In the B band, the differences are more remarkable: Prior to lockdown: 18.77 mag/arcsec$^2$. During lockdown 19.28 mag/arcsec$^2$. This means that approximately a 45$\%$ reduction in the light output of the city is derived from our data.

Note that private lighting is a significant contributor to light pollution \citep{kyba2020}, and is usually bluer than municipal lighting, at least in some cities where cold led lighting is not extended yet. This indicates that the blue band night sky brightness appears to be a better human activity indicator than the sky brightness in the visible channel. Even though the outdoor human activities dropped by more than 90$\%$, the light output of the city did not decrease by that amount, clearly pointing out that most of the city lighting does not adapt to the real use of its citizens. This shows why measurements from VIIRS are not optimal to trace the pandemic effects on the economy because of the flyby time of the satellite. An instrument with an earlier flyby time is urgently needed to track changes in urban light emissions.

\section{Conclusions}

We found a clear decrease in light pollution (as measured in terms of sky brightness) within the city of Granada during the lockdown. Also, the sky glow from Granada as seen at a particular angle of 25 degrees outside the city, from Sierra Nevada observatory, decreased. The observations are consistent with the idea that the decrease arose for two reasons. First, this was because of a decrease in scattered light due to the presence of fewer anthropogenic aerosols during the lockdown. In this regard, a clear correlation is found between sky brightness at several wavelength bands and PM10 particle concentration measured within the city. Outside the city at 25 degrees, the correlation coefficient with the PM10 particle concentration in Granada is lower, which makes sense given that the aerosols that contribute are not constrained to the downtown area. The other main reason for the reduction in the light pollution comes from a decrease in the net amount of light emitted by the city at the level of 20$\%$ in the visible and $45\%$ in the B band, probably due to a decrease in private lighting of buildings, vehicle lights and lighting of private areas, mainly observable at the start of the night. Very late at night, the flux from the city during lockdown and prior to it was almost unchanged. Satellite imagery does not show marked differences during the lockdown in terms of total light output (at the 10$\%$ precision level) because these images were obtained late at night, consistent with the ground based data, which show little change in the second half of the night. Given that outdoor activities decreased by up to 90$\%$ during lockdown, our results indicate that the light output of the city seems to be dominated by permanent lighting that does not adapt to the real use of the outdoor areas by the citizens (at least in the case of Granada), pointing out a clear waste of energy and resources.

\section{Acknowledgements}
This research was partially based on data taken at the Sierra Nevada Observatory, which is operated by the Instituto de Astrof\'isica de Andaluc\'ia (IAA-CSIC). Funding from Spanish project AYA2017-89637-R, and from FEDER, is acknowledged. We also acknowledge financial support from the State Agency for Research of the Spanish MCIU through the "Center of Excellence Severo Ochoa" award for the Instituto de Astrof\'isica de Andaluc\'ia (SEV- 2017-0709).The authors thank IAA-CSIC
Computer Center and Maintenance Staff for their help in the installation of the SQM devices. 
This work was supported by the EMISSI@N project (NERC grant NE/P01156X/1).

\section*{Description of author's responsibilities}
M.B.C., A.S.M., S.M.R., J.L.O., J.M.V., A.P., J.Z., J.B. and K.J.G. conceived the study, M.B.C., A.S.M., S.M.R., J.L.O., A.G. and J. Z.  conducted the calibration procedures and acquisition, M.B.C., S.M.R., A.P  made the management, permits and data request, M.B.C., A.S.M., S.M.R., J.L.O., J.M.V., J.B. and K.J.G. wrote the original manuscript, S.M.R., J.L.O., A.S.M, J.Z., J.M.V., J.B. and K.J.G. conducted the funding requests. All authors reviewed the manuscript.

\bibliography{sample}

\begin{thebibliography}{37}
\expandafter\ifx\csname natexlab\endcsname\relax\def\natexlab#1{#1}\fi
\providecommand{\url}[1]{\texttt{#1}}
\providecommand{\href}[2]{#2}
\providecommand{\path}[1]{#1}
\providecommand{\DOIprefix}{doi:}
\providecommand{\ArXivprefix}{arXiv:}
\providecommand{\URLprefix}{URL: }
\providecommand{\Pubmedprefix}{pmid:}
\providecommand{\doi}[1]{\href{http://dx.doi.org/#1}{\path{#1}}}
\providecommand{\Pubmed}[1]{\href{pmid:#1}{\path{#1}}}
\providecommand{\bibinfo}[2]{#2}
\ifx\xfnm\relax \def\xfnm[#1]{\unskip,\space#1}\fi
%Type = Article
\bibitem[{{Aceituno} et~al.(2011){Aceituno}, {S{\'a}nchez}, {Aceituno},
  {Galad{\'\i}-Enr{\'\i}quez}, {Negro}, {Soriguer} and {Sanchez
  Gomez}}]{aceituno2011}
\bibinfo{author}{{Aceituno}, J.}, \bibinfo{author}{{S{\'a}nchez}, S.F.},
  \bibinfo{author}{{Aceituno}, F.J.},
  \bibinfo{author}{{Galad{\'\i}-Enr{\'\i}quez}, D.}, \bibinfo{author}{{Negro},
  J.J.}, \bibinfo{author}{{Soriguer}, R.C.}, \bibinfo{author}{{Sanchez Gomez},
  G.}, \bibinfo{year}{2011}.
\newblock \bibinfo{title}{An all-sky transmission monitor: Astmon}.
\newblock \bibinfo{journal}{Publications of the Astronomical Society of the
  Pacific} \bibinfo{volume}{123}, \bibinfo{pages}{1076--1086}.
%Type = Misc
\bibitem[{AEMET(2020)}]{AEMET}
\bibinfo{author}{AEMET}, \bibinfo{year}{2020}.
\newblock \bibinfo{title}{Informe mensual climatológico}.
\newblock \URLprefix
  \url{http://www.aemet.es/documentos/es/serviciosclimaticos/vigilancia_clima/resumenes_climat/mensuales/2020/res_mens_clim_2020_04.pdf}.
%Type = Misc
\bibitem[{Ayto.Granada(2020)}]{calidadaire}
\bibinfo{author}{Ayto.Granada}, \bibinfo{year}{2020}.
\newblock \bibinfo{title}{Calidad del aire}.
\newblock \URLprefix
  \url{https://www.granada.org/inet/calidadaire.nsf/icayear}.
%Type = Article
\bibitem[{Bar{\'a} et~al.(2019)Bar{\'a}, Rodr{\'\i}guez-Ar{\'o}s, P{\'e}rez,
  Tosar, Lima, S{\'a}nchez~de Miguel and Zamorano}]{bara2019estimating}
\bibinfo{author}{Bar{\'a}, S.}, \bibinfo{author}{Rodr{\'\i}guez-Ar{\'o}s,
  {\'A}.}, \bibinfo{author}{P{\'e}rez, M.}, \bibinfo{author}{Tosar, B.},
  \bibinfo{author}{Lima, R.C.}, \bibinfo{author}{S{\'a}nchez~de Miguel, A.},
  \bibinfo{author}{Zamorano, J.}, \bibinfo{year}{2019}.
\newblock \bibinfo{title}{Estimating the relative contribution of streetlights,
  vehicles, and residential lighting to the urban night sky brightness}.
\newblock \bibinfo{journal}{Lighting Research \& Technology}
  \bibinfo{volume}{51}, \bibinfo{pages}{1092--1107}.
%Type = Misc
\bibitem[{Bui and Badger(2020)}]{nyt}
\bibinfo{author}{Bui, Q.}, \bibinfo{author}{Badger, E.}, \bibinfo{year}{2020}.
\newblock \bibinfo{title}{The coronavirus quieted city noise. listen to
  what’s left.}
\newblock \URLprefix
  \url{https://www.nytimes.com/interactive/2020/05/22/upshot/coronavirus-quiet-city-noise.html}.
%Type = Article
\bibitem[{Cheung et~al.(2015)Cheung, Pun, SO, Shibata, Walker and
  Agata}]{cheung2015globe}
\bibinfo{author}{Cheung, S.L.}, \bibinfo{author}{Pun, J.C.S.},
  \bibinfo{author}{SO, C.w.}, \bibinfo{author}{Shibata, Y.},
  \bibinfo{author}{Walker, C.E.}, \bibinfo{author}{Agata, H.},
  \bibinfo{year}{2015}.
\newblock \bibinfo{title}{Globe at night-sky brightness monitoring network}.
\newblock \bibinfo{journal}{IAUGA} \bibinfo{volume}{29},
  \bibinfo{pages}{2257516}.
%Type = Article
\bibitem[{Cinzano(2005)}]{cinzano2005}
\bibinfo{author}{Cinzano, P.}, \bibinfo{year}{2005}.
\newblock \bibinfo{title}{Night sky photometry with sky quality meter}.
\newblock \bibinfo{journal}{Internal Report, ISTIL} \bibinfo{volume}{9},
  \bibinfo{pages}{1--13}.
\newblock \bibinfo{note}{Available at:
  http://www.lightpollution.it/download/sqmreport.pdf}.
%Type = Article
\bibitem[{Elvidge et~al.(2020)Elvidge, Ghosh, Hsu, Zhizhin and
  Bazilian}]{elvidge2020dimming}
\bibinfo{author}{Elvidge, C.D.}, \bibinfo{author}{Ghosh, T.},
  \bibinfo{author}{Hsu, F.C.}, \bibinfo{author}{Zhizhin, M.},
  \bibinfo{author}{Bazilian, M.}, \bibinfo{year}{2020}.
\newblock \bibinfo{title}{The dimming of lights in china during the covid-19
  pandemic}.
\newblock \bibinfo{journal}{Remote Sensing} \bibinfo{volume}{12},
  \bibinfo{pages}{2851}.
%Type = Article
\bibitem[{Falchi et~al.(2016)Falchi, Cinzano, Duriscoe, Kyba, Elvidge, Baugh,
  Portnov, Rybnikova and Furgoni}]{falchi2016new}
\bibinfo{author}{Falchi, F.}, \bibinfo{author}{Cinzano, P.},
  \bibinfo{author}{Duriscoe, D.}, \bibinfo{author}{Kyba, C.C.},
  \bibinfo{author}{Elvidge, C.D.}, \bibinfo{author}{Baugh, K.},
  \bibinfo{author}{Portnov, B.A.}, \bibinfo{author}{Rybnikova, N.A.},
  \bibinfo{author}{Furgoni, R.}, \bibinfo{year}{2016}.
\newblock \bibinfo{title}{The new world atlas of artificial night sky
  brightness}.
\newblock \bibinfo{journal}{Science Advances} \bibinfo{volume}{2},
  \bibinfo{pages}{e1600377}.
%Type = Article
\bibitem[{Falchi et~al.(2011)Falchi, Cinzano, Elvidge, Keith and
  Haim}]{falchi2011limiting}
\bibinfo{author}{Falchi, F.}, \bibinfo{author}{Cinzano, P.},
  \bibinfo{author}{Elvidge, C.D.}, \bibinfo{author}{Keith, D.M.},
  \bibinfo{author}{Haim, A.}, \bibinfo{year}{2011}.
\newblock \bibinfo{title}{Limiting the impact of light pollution on human
  health, environment and stellar visibility}.
\newblock \bibinfo{journal}{Journal of environmental management}
  \bibinfo{volume}{92}, \bibinfo{pages}{2714--2722}.
%Type = Article
\bibitem[{Garstang(1986)}]{garstang1986model}
\bibinfo{author}{Garstang, R.}, \bibinfo{year}{1986}.
\newblock \bibinfo{title}{Model for artificial night-sky illumination}.
\newblock \bibinfo{journal}{Publications of the Astronomical Society of the
  Pacific} \bibinfo{volume}{98}, \bibinfo{pages}{364--375}.
%Type = Article
\bibitem[{Garstang(1991)}]{garstang1991modeling}
\bibinfo{author}{Garstang, R.}, \bibinfo{year}{1991}.
\newblock \bibinfo{title}{Light pollution modeling}.
\newblock \bibinfo{journal}{Publications of the Astronomical Society of the
  Pacific} \bibinfo{volume}{17}, \bibinfo{pages}{56--69}.
%Type = Article
\bibitem[{Gaston et~al.(2013)Gaston, Bennie, Davies and
  Hopkins}]{gaston2013ecological}
\bibinfo{author}{Gaston, K.J.}, \bibinfo{author}{Bennie, J.},
  \bibinfo{author}{Davies, T.W.}, \bibinfo{author}{Hopkins, J.},
  \bibinfo{year}{2013}.
\newblock \bibinfo{title}{The ecological impacts of nighttime light pollution:
  a mechanistic appraisal}.
\newblock \bibinfo{journal}{Biological reviews} \bibinfo{volume}{88},
  \bibinfo{pages}{912--927}.
%Type = Article
\bibitem[{Ghosh et~al.(2020)Ghosh, Elvidge, Hsu, Zhizhin and
  Bazilian}]{ghosh2020dimming}
\bibinfo{author}{Ghosh, T.}, \bibinfo{author}{Elvidge, C.D.},
  \bibinfo{author}{Hsu, F.C.}, \bibinfo{author}{Zhizhin, M.},
  \bibinfo{author}{Bazilian, M.}, \bibinfo{year}{2020}.
\newblock \bibinfo{title}{The dimming of lights in india during the covid-19
  pandemic}.
\newblock \bibinfo{journal}{Remote Sensing} \bibinfo{volume}{12},
  \bibinfo{pages}{3289}.
%Type = Article
\bibitem[{H{\"a}nel et~al.(2018)H{\"a}nel, Posch, Ribas, Aub{\'e}, Duriscoe,
  Jechow, Kollath, Lolkema, Moore, Schmidt, {Spoelstra}, {Wuchterl} and
  {Kyba}}]{hanel2018}
\bibinfo{author}{H{\"a}nel, A.}, \bibinfo{author}{Posch, T.},
  \bibinfo{author}{Ribas, S.J.}, \bibinfo{author}{Aub{\'e}, M.},
  \bibinfo{author}{Duriscoe, D.}, \bibinfo{author}{Jechow, A.},
  \bibinfo{author}{Kollath, Z.}, \bibinfo{author}{Lolkema, D.E.},
  \bibinfo{author}{Moore, C.}, \bibinfo{author}{Schmidt, N.},
  \bibinfo{author}{{Spoelstra}, H.}, \bibinfo{author}{{Wuchterl}, G.},
  \bibinfo{author}{{Kyba}, C.C.M.}, \bibinfo{year}{2018}.
\newblock \bibinfo{title}{Measuring night sky brightness: methods and
  challenges}.
\newblock \bibinfo{journal}{Journal of Quantitative Spectroscopy and Radiative
  Transfer} \bibinfo{volume}{205}, \bibinfo{pages}{278--290}.
%Type = Article
\bibitem[{Jechow and H\"olker(2020)}]{Jechow2020Berlin}
\bibinfo{author}{Jechow, A.}, \bibinfo{author}{H\"olker, F.},
  \bibinfo{year}{2020}.
\newblock \bibinfo{title}{Evidence that reduced air and road traffic decreased
  artificial night-time skyglow during covid-19 lockdown in berlin, germany}.
\newblock \bibinfo{journal}{Remote Sensing} \bibinfo{volume}{12},
  \bibinfo{pages}{3412}.
%Type = Inproceedings
\bibitem[{Joye and Mandel(2003)}]{joye2003new}
\bibinfo{author}{Joye, W.}, \bibinfo{author}{Mandel, E.}, \bibinfo{year}{2003}.
\newblock \bibinfo{title}{New features of saoimage ds9}, in:
  \bibinfo{booktitle}{Astronomical data analysis software and systems XII}, p.
  \bibinfo{pages}{489}.
%Type = Article
\bibitem[{Koo et~al.(2016)Koo, Song, Joo, Lee, Lee, Lee and
  Jung}]{koo2016outdoor}
\bibinfo{author}{Koo, Y.S.}, \bibinfo{author}{Song, J.Y.},
  \bibinfo{author}{Joo, E.Y.}, \bibinfo{author}{Lee, H.J.},
  \bibinfo{author}{Lee, E.}, \bibinfo{author}{Lee, S.k.},
  \bibinfo{author}{Jung, K.Y.}, \bibinfo{year}{2016}.
\newblock \bibinfo{title}{Outdoor artificial light at night, obesity, and sleep
  health: Cross-sectional analysis in the koges study}.
\newblock \bibinfo{journal}{Chronobiology international} \bibinfo{volume}{33},
  \bibinfo{pages}{301--314}.
%Type = Article
\bibitem[{Kyba et~al.(2020)Kyba, Ruby, Kuechly, Kinzey, Miller, Sanders,
  Barentine, Kleinodt and Espey}]{kyba2020}
\bibinfo{author}{Kyba, C.}, \bibinfo{author}{Ruby, A.},
  \bibinfo{author}{Kuechly, H.}, \bibinfo{author}{Kinzey, B.},
  \bibinfo{author}{Miller, N.}, \bibinfo{author}{Sanders, J.},
  \bibinfo{author}{Barentine, J.}, \bibinfo{author}{Kleinodt, R.},
  \bibinfo{author}{Espey, B.}, \bibinfo{year}{2020}.
\newblock \bibinfo{title}{Direct measurement of the contribution of street
  lighting to satellite observations of nighttime light emissions from urban
  areas}.
\newblock \bibinfo{journal}{Lighting Research and Technology}
  \bibinfo{note}{Avaiable at: https://doi.org/10.1177/1477153520958463}.
%Type = Article
\bibitem[{Kyba et~al.(2012)Kyba, Ruhtz, Fischer and H{\"o}lker}]{kyba2012red}
\bibinfo{author}{Kyba, C.}, \bibinfo{author}{Ruhtz, T.},
  \bibinfo{author}{Fischer, J.}, \bibinfo{author}{H{\"o}lker, F.},
  \bibinfo{year}{2012}.
\newblock \bibinfo{title}{Red is the new black: how the colour of urban skyglow
  varies with cloud cover}.
\newblock \bibinfo{journal}{Monthly Notices of the Royal Astronomical Society}
  \bibinfo{volume}{425}, \bibinfo{pages}{701--708}.
%Type = Article
\bibitem[{{Kyba} et~al.(2015){Kyba}, {Tong}, {Bennie}, {Birriel}, {Birriel},
  {Cool}, {Danielsen}, {Davies}, {Outer}, {Edwards}, {Ehlert}, {Falchi},
  {Fischer}, {Giacomelli}, {Giubbilini}, {Haaima}, {Hesse}, {Heygster},
  {H{\"o}lker}, {Inger}, {Jensen}, {Kuechly}, {Kuehn}, {Langill}, {Lolkema},
  {Nagy}, {Nievas}, {Ochi}, {Popow}, {Posch}, {Puschnig}, {Ruhtz}, {Schmidt},
  {Schwarz}, {Schwope}, {Spoelstra}, {Tekatch}, {Trueblood}, {Walker}, {Weber},
  {Welch}, {Zamorano} and {Gaston}}]{kyba2015}
\bibinfo{author}{{Kyba}, C.}, \bibinfo{author}{{Tong}, K.P.},
  \bibinfo{author}{{Bennie}, J.}, \bibinfo{author}{{Birriel}, I.},
  \bibinfo{author}{{Birriel}, J.J.}, \bibinfo{author}{{Cool}, A.},
  \bibinfo{author}{{Danielsen}, A.}, \bibinfo{author}{{Davies}, T.W.},
  \bibinfo{author}{{Outer}, P.N.D.}, \bibinfo{author}{{Edwards}, W.},
  \bibinfo{author}{{Ehlert}, R.}, \bibinfo{author}{{Falchi}, F.},
  \bibinfo{author}{{Fischer}, J.}, \bibinfo{author}{{Giacomelli}, A.},
  \bibinfo{author}{{Giubbilini}, F.}, \bibinfo{author}{{Haaima}, M.},
  \bibinfo{author}{{Hesse}, C.}, \bibinfo{author}{{Heygster}, G.},
  \bibinfo{author}{{H{\"o}lker}, F.}, \bibinfo{author}{{Inger}, R.},
  \bibinfo{author}{{Jensen}, L.J.}, \bibinfo{author}{{Kuechly}, H.U.},
  \bibinfo{author}{{Kuehn}, J.}, \bibinfo{author}{{Langill}, P.},
  \bibinfo{author}{{Lolkema}, D.E.}, \bibinfo{author}{{Nagy}, M.},
  \bibinfo{author}{{Nievas}, M.}, \bibinfo{author}{{Ochi}, N.},
  \bibinfo{author}{{Popow}, E.}, \bibinfo{author}{{Posch}, T.},
  \bibinfo{author}{{Puschnig}, J.}, \bibinfo{author}{{Ruhtz}, T.},
  \bibinfo{author}{{Schmidt}, W.}, \bibinfo{author}{{Schwarz}, R.},
  \bibinfo{author}{{Schwope}, A.}, \bibinfo{author}{{Spoelstra}, H.},
  \bibinfo{author}{{Tekatch}, A.}, \bibinfo{author}{{Trueblood}, M.},
  \bibinfo{author}{{Walker}, C.E.}, \bibinfo{author}{{Weber}, M.},
  \bibinfo{author}{{Welch}, D.L.}, \bibinfo{author}{{Zamorano}, J.},
  \bibinfo{author}{{Gaston}, K.J.}, \bibinfo{year}{2015}.
\newblock \bibinfo{title}{Worldwide variations in artificial skyglow}.
\newblock \bibinfo{journal}{Scientific Reports} \bibinfo{volume}{5},
  \bibinfo{pages}{1--6}.
%Type = Article
\bibitem[{Levin et~al.(2020)Levin, Kyba, Zhang, de~Miguel, Rom{\'a}n, Li,
  Portnov, Molthan, Jechow, Miller et~al.}]{levin2020remote}
\bibinfo{author}{Levin, N.}, \bibinfo{author}{Kyba, C.C.},
  \bibinfo{author}{Zhang, Q.}, \bibinfo{author}{de~Miguel, A.S.},
  \bibinfo{author}{Rom{\'a}n, M.O.}, \bibinfo{author}{Li, X.},
  \bibinfo{author}{Portnov, B.A.}, \bibinfo{author}{Molthan, A.L.},
  \bibinfo{author}{Jechow, A.}, \bibinfo{author}{Miller, S.D.}, et~al.,
  \bibinfo{year}{2020}.
\newblock \bibinfo{title}{Remote sensing of night lights: A review and an
  outlook for the future}.
\newblock \bibinfo{journal}{Remote Sensing of Environment}
  \bibinfo{volume}{237}, \bibinfo{pages}{111443}.
%Type = Article
\bibitem[{Levin and Zhang(2017)}]{levin2017global}
\bibinfo{author}{Levin, N.}, \bibinfo{author}{Zhang, Q.}, \bibinfo{year}{2017}.
\newblock \bibinfo{title}{A global analysis of factors controlling viirs
  nighttime light levels from densely populated areas}.
\newblock \bibinfo{journal}{Remote sensing of environment}
  \bibinfo{volume}{190}, \bibinfo{pages}{366--382}.
%Type = Misc
\bibitem[{LightPollutionMap(2020)}]{lpm}
\bibinfo{author}{LightPollutionMap}, \bibinfo{year}{2020}.
\newblock \bibinfo{title}{Light pollution map}.
\newblock \URLprefix \url{https://www.lightpollutionmap.info/}.
%Type = Article
\bibitem[{Liu et~al.(2020)Liu, Sha, Liu, Houser, Zhang, Hou, Lan, Flynn, Lu, Hu
  et~al.}]{liu2020spatiotemporal}
\bibinfo{author}{Liu, Q.}, \bibinfo{author}{Sha, D.}, \bibinfo{author}{Liu,
  W.}, \bibinfo{author}{Houser, P.}, \bibinfo{author}{Zhang, L.},
  \bibinfo{author}{Hou, R.}, \bibinfo{author}{Lan, H.}, \bibinfo{author}{Flynn,
  C.}, \bibinfo{author}{Lu, M.}, \bibinfo{author}{Hu, T.}, et~al.,
  \bibinfo{year}{2020}.
\newblock \bibinfo{title}{Spatiotemporal patterns of covid-19 impact on human
  activities and environment in mainland china using nighttime light and air
  quality data}.
\newblock \bibinfo{journal}{Remote Sensing} \bibinfo{volume}{12},
  \bibinfo{pages}{1576}.
%Type = Article
\bibitem[{Mandel et~al.(2011)Mandel, Murray and Roll}]{mandel2011funtools}
\bibinfo{author}{Mandel, E.}, \bibinfo{author}{Murray, S.S.},
  \bibinfo{author}{Roll, J.}, \bibinfo{year}{2011}.
\newblock \bibinfo{title}{Funtools: Fits users need tools}.
\newblock \bibinfo{journal}{ascl} , \bibinfo{pages}{ascl--1112}.
%Type = Misc
\bibitem[{NASA(2020a)}]{nasaearth}
\bibinfo{author}{NASA}, \bibinfo{year}{2020}a.
\newblock \bibinfo{title}{Airborne particle levels plummet in northern india}.
\newblock \URLprefix
  \url{https://earthobservatory.nasa.gov/images/146596/airborne-particle-levels-plummet-in-northern-india?src=nha}.
%Type = Misc
\bibitem[{NASA(2020b)}]{nasaair}
\bibinfo{author}{NASA}, \bibinfo{year}{2020}b.
\newblock \bibinfo{title}{Nasa satellite data show 30 percent drop in air
  pollution over northeast u.s.}
\newblock \URLprefix
  \url{https://www.nasa.gov/feature/goddard/2020/drop-in-air-pollution-over-northeast}.
%Type = Article
\bibitem[{Nievas~Rosillo and Zamorano(2014)}]{nievas2014}
\bibinfo{author}{Nievas~Rosillo, M.}, \bibinfo{author}{Zamorano, J.},
  \bibinfo{year}{2014}.
\newblock \bibinfo{title}{Pysqm the ucm open source software to read, plot and
  store data from sqm photometers}.
\newblock \bibinfo{journal}{Internal Report, UCM} ,
  \bibinfo{pages}{1--8}\bibinfo{note}{Available at:
  https://eprints.ucm.es/25900/1/LICA\_PySQM\_v2.pdf}.
%Type = Book
\bibitem[{Rich and Longcore(2013)}]{rich2013ecological}
\bibinfo{author}{Rich, C.}, \bibinfo{author}{Longcore, T.},
  \bibinfo{year}{2013}.
\newblock \bibinfo{title}{Ecological consequences of artificial night
  lighting}.
\newblock \bibinfo{publisher}{Island Press}.
%Type = Article
\bibitem[{Rom{\'a}n et~al.(2018)Rom{\'a}n, Wang, Sun, Kalb, Miller, Molthan,
  Schultz, Bell, Stokes, Pandey et~al.}]{roman2018nasa}
\bibinfo{author}{Rom{\'a}n, M.O.}, \bibinfo{author}{Wang, Z.},
  \bibinfo{author}{Sun, Q.}, \bibinfo{author}{Kalb, V.},
  \bibinfo{author}{Miller, S.D.}, \bibinfo{author}{Molthan, A.},
  \bibinfo{author}{Schultz, L.}, \bibinfo{author}{Bell, J.},
  \bibinfo{author}{Stokes, E.C.}, \bibinfo{author}{Pandey, B.}, et~al.,
  \bibinfo{year}{2018}.
\newblock \bibinfo{title}{Nasa's black marble nighttime lights product suite}.
\newblock \bibinfo{journal}{Remote Sensing of Environment}
  \bibinfo{volume}{210}, \bibinfo{pages}{113--143}.
%Type = Phdthesis
\bibitem[{{S\'anchez de Miguel}(2015)}]{de2015variacion}
\bibinfo{author}{{S\'anchez de Miguel}, A.}, \bibinfo{year}{2015}.
\newblock \bibinfo{title}{Variacion espacial, temporal y espectral de la
  contaminacion lum{\i}nica y sus fuentes: Metodolog{\i}a y resultados}.
\newblock Ph.D. thesis. Universidad Complutense de Madrid.
\newblock \DOIprefix\doi{10.5281/zenodo.1289932}.
%Type = Article
\bibitem[{Small and Elvidge(2013)}]{small2013night}
\bibinfo{author}{Small, C.}, \bibinfo{author}{Elvidge, C.D.},
  \bibinfo{year}{2013}.
\newblock \bibinfo{title}{Night on earth: Mapping decadal changes of
  anthropogenic night light in asia}.
\newblock \bibinfo{journal}{International Journal of Applied Earth Observation
  and Geoinformation} \bibinfo{volume}{22}, \bibinfo{pages}{40--52}.
%Type = Article
\bibitem[{{Sánchez de Miguel} et~al.(2017){Sánchez de Miguel}, Aubé,
  Zamorano, Kocifaj, Roby and Tapia}]{doi:10.1093/mnras/stx145}
\bibinfo{author}{{Sánchez de Miguel}, A.}, \bibinfo{author}{Aubé, M.},
  \bibinfo{author}{Zamorano, J.}, \bibinfo{author}{Kocifaj, M.},
  \bibinfo{author}{Roby, J.}, \bibinfo{author}{Tapia, C.},
  \bibinfo{year}{2017}.
\newblock \bibinfo{title}{Sky quality meter measurements in a colour-changing
  world}.
\newblock \bibinfo{journal}{Monthly Notices of the Royal Astronomical Society}
  \bibinfo{volume}{467}, \bibinfo{pages}{2966}.
%Type = Inproceedings
\bibitem[{Walker et~al.(2008)Walker, Pompea and Isbell}]{walker2008globe}
\bibinfo{author}{Walker, C.E.}, \bibinfo{author}{Pompea, S.M.},
  \bibinfo{author}{Isbell, D.}, \bibinfo{year}{2008}.
\newblock \bibinfo{title}{Globe at night 2.0: On the road toward iya 2009}, in:
  \bibinfo{booktitle}{EPO and a Changing World: Creating Linkages and Expanding
  Partnerships}, p. \bibinfo{pages}{423}.
%Type = Misc
\bibitem[{Xiao et~al.(2020)Xiao, Eilon, Ji and Tanimoto}]{xiao2020covid19}
\bibinfo{author}{Xiao, H.}, \bibinfo{author}{Eilon, Z.}, \bibinfo{author}{Ji,
  C.}, \bibinfo{author}{Tanimoto, T.}, \bibinfo{year}{2020}.
\newblock \bibinfo{title}{Covid-19 societal response captured by seismic noise
  in china and italy}.
\newblock \href{http://arxiv.org/abs/2005.00131}{\tt arXiv:2005.00131}.
%Type = Article
\bibitem[{Zamorano et~al.(2016)Zamorano, Garc{\'\i}a, Tapia, de~Miguel, Pascual
  and Gallego}]{zamorano2016stars4all}
\bibinfo{author}{Zamorano, J.}, \bibinfo{author}{Garc{\'\i}a, C.},
  \bibinfo{author}{Tapia, C.}, \bibinfo{author}{de~Miguel, A.S.},
  \bibinfo{author}{Pascual, S.}, \bibinfo{author}{Gallego, J.},
  \bibinfo{year}{2016}.
\newblock \bibinfo{title}{Stars4all night sky brightness photometer}.
\newblock \bibinfo{journal}{International Journal of Sustainable Lighting}
  \bibinfo{volume}{18}, \bibinfo{pages}{49--54}.

\end{thebibliography}

\appendix
\clearpage
\appendixpage
\section{Measurements of the observatory of La Sagra}\label{A:1}
The observatory of La Sagra is located in one of the darkest areas of Spain, so the low impact of light pollution allows the use of measurements taken from its location as a reference to identify the natural variations in sky brightness at the zenith between different time intervals and months. For this purpose, we have studied the data collected since 2018 by a SQM device (without added filter), selecting the valid nights of the months of February, April and May in the same time bands as for the measurements in Granada.

Figure A.9 shows the average brightness of the sky from 21:00 to 22:00 and from 2:00 to 3:00 UT for a total of 30 nights (14 distributed in February and 16 in April and May). During February, in the time band before midnight the sky at the zenith is 0.28 $mag/arcsec^{2}$ brighter than in the early morning hours. In spring the situation is the opposite (although the difference is very small): from 2:00 to 3:00 is 0.05 $mag/arcsec^{2}$ brighter than from 21:00 to 22:00 UT. If we compare the same time range, we see that from 21:00 to 22:00 UT the sky is 0.27 $mag/arcsec^{2}$ darker in the spring months while from 3:00 to 4:00 UT it is 0.06 $mag/arcsec^{2}$ brighter than in February. 

\begin{figure}[ht]
\begin{center}
\includegraphics[scale=0.25]{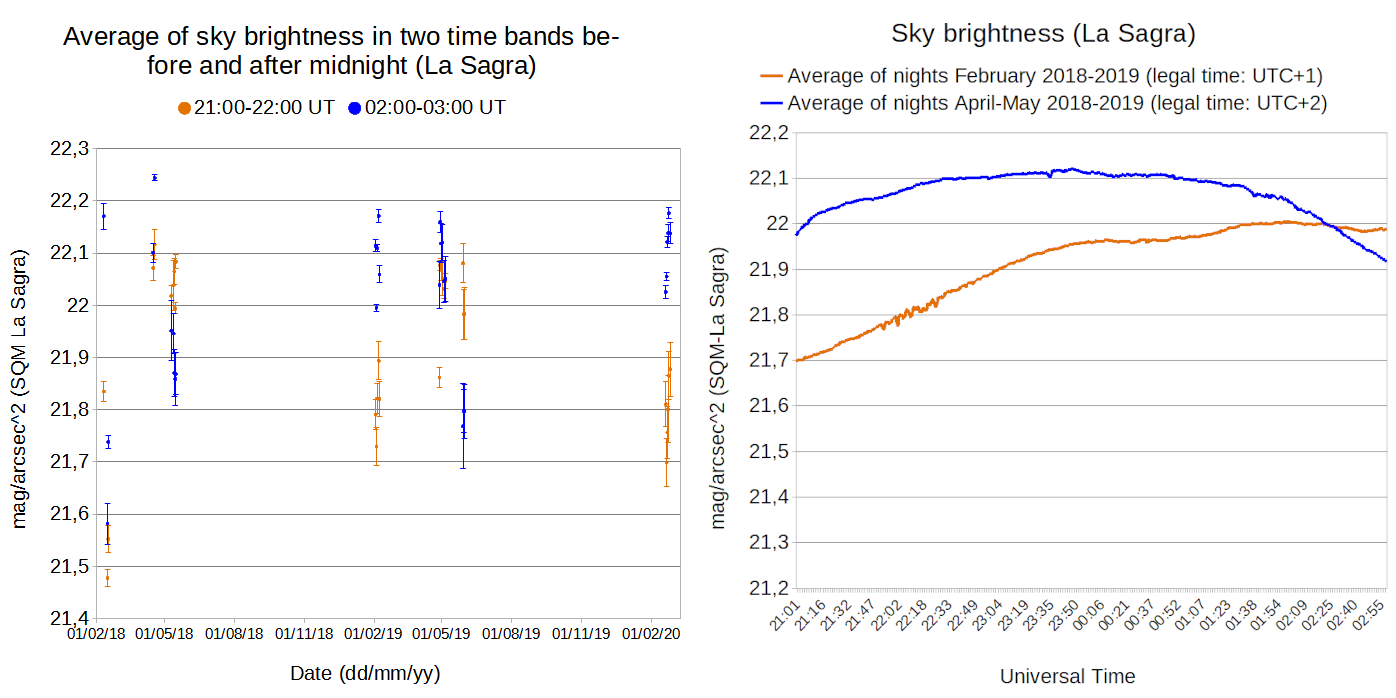}
\caption{Left: time-averaged sky brightness from the unfiltered SQM photometer as a function of date. Right: night-averaged sky brightness from the unfiltered SQM photometer as a function of time. Photometers located at La Sagra Observatory.}
\end{center}
\label{fig:SQM5}
\end{figure}

It is clear that there is a natural variation of sky brightness at the zenith due to the transit of different regions with different densities of stars. But to what extent are these variations perceptible in a urban environment? In order to understand this, it is necessary to put these differences into context, since in La Sagra they occur with values between 21.70 and 22 $ mag/arcsec^{2} $, in a sky of exceptional darkness, while in the city of Granada there is an artificial brightness which means that even in the best conditions the 19 $ mag/arcsec^{2} $ are not reached.  In terms of light intensity per unit area ($ mcd/m^{2} $), a difference of 0.30 $ mag/arcsec^{2} $ in La Sagra is approximately 0.061 $ mcd/m^{2} $, while a difference of 0.30 $ mag/arcsec^{2} $ in the city of Granada is 1 $ mcd/m^{2} $. In other words, in terms of luminance, the hourly or seasonal variation of the natural brightness of the sky cannot account for more than 6.1\% of the differences observed in the city of Granada \citep{falchi2016new} taken from \cite{lpm} web. 

Similarly, the hourly evolution of the average of the records of sky brightness at La Sagra for February and the months of April and May (2018-2019) can be taken as a reference of the natural variation due to the transit of different star fields through the zenith. But this is not significant in the case of the urban measurements obtained from IAA-CSIC, as the higher luminance of artificial origin masks the natural variations by almost 95\%.
\clearpage
\section{Spearman's correlation coefficients.}\label{A:2}
\begin{table}[ht]
\begin{center}
\caption{Spearman's correlation coefficients for sky brightness measured from IAA in Granada, air pollution and time.\textit{Time} is the time measured as a fraction of a Julian day; \textit{PM10} and \textit{NO2} are the particle concentrations (between 2.5 and 10 microns) and of nitrogen dioxide ($ \mu g/m^{3} $), \textit{SQM}, \textit{SQMB} and \textit{SQMV} are the values of sky brightness without filter, with filter B and with filter V ($ mag/arcsec^{2} $) and \textit{B.V} is the color index B-V, as a result of subtracting the measurement of \textit{SQMB} and \textit{SQMV}.\newline}
\scalebox{0.7}{\begin{tabular}{||c|c|c|c|c|c|c|c||}
\hline 
\hline 
 & Time & PM10 & NO2 & SQM & SQMB & SQMV & B.V \\ 
\hline 
Time & 1 & -0.65 & -0.62 & 0.83 & 0.85 & 0.8 & 0.65 \\ 
\hline 
PM10 & -0.65 & 1 & 0.69 & -0.84 & -0.77 & -0.84 & -0.44 \\ 
\hline 
NO2 & -0.62 & 0.69 & 1 & -0.69 & -0.81 & -0.68 & -0.77 \\ 
\hline 
SQM & 0.83 & -0.84 & -0.69 & 1 & 0.92 & 1 & 0.53 \\ 
\hline 
SQMB & 0.85 & -0.77 & -0.81 & 0.92 & 1 & 0.91 & 0.78 \\ 
\hline 
SQMV & 0.8 & -0.84 & -0.68 & 1 & 0.91 & 1 & 0.51 \\ 
\hline 
B.V & 0.65 & -0.44 & -0.77 & 0.53 & 0.78 & 0.51 & 1 \\ 
\hline
\hline 
\end{tabular}}
\end{center}
\end{table}

\begin{table}[ht]
\begin{center}
\caption{Spearman's correlation coefficients for sky brightness measured from OSN at $\sim$25º above Granada, air pollution in Granada and time.\textit{ASTMON.V}, \textit{ASTMON.B} and \textit{ASTMON.R} are the values of sky brightness (with different Johnson filters) of the ASTMON device installed at Sierra Nevada Observatory.\newline}
\scalebox{0.7}{\begin{tabular}{||c|c|c|c|c|c|c||}
\hline 
\hline 
 & Time & PM10 & NO2 & \tiny ASTMON.V & \tiny ASTMON.B & \tiny ASTMON.R \\ 
\hline 
Time & 1 & -0.61 & -0.54 & 0.76 & 0.83 & 0.73 \\ 
\hline 
PM10 & -0.61 & 1 & 0.62 & -0.6 & -0.64 & -0.57 \\ 
\hline 
NO2 & -0.54 & 0.62 & 1 & -0.54 & -0.66 & -0.51 \\ 
\hline 
\tiny ASTMON.V & 0.76 & -0.6 & -0.54 & 1 & 0.8 & 0.95 \\ 
\hline 
\tiny ASTMON.B & 0.83 & -0.64 & -0.66 & 0.8 & 1 & 0.76 \\ 
\hline 
\tiny ASTMON.R & 0.73 & -0.57 & -0.51 & 0.95 & 0.76 & 1 \\ 
\hline
\hline 
\end{tabular}}
\end{center}
\end{table}
\clearpage

\section{Fits of lineal models.}\label{A:3}

\begin{table}[ht]
\begin{center}
\caption{Fit of lineal model between sky brightness (SQM without filter) and PM10 particle concentration.\newline}
\scalebox{0.80}{\begin{tabular}{ ||m{2cm} m{1.5cm} m{1.5cm} m{1.5cm} m{1.5cm}|| }
\hline
\hline
\small Coefficients: & \small Estimate & \small Std. Error & \small t value & \small $ Pr(>|t|) $ \\
\hline
\hline
\small (Intercept) & 18.908003 & 0.024691 & 765.78 & $ < 2e-16 $ \\
PM10 & -0.017914 & 0.000801 & -22.36 & $ < 2e-16 $ \\
\hline
\hline
\multicolumn{5}{||m{10cm}||}{Residuals:} \\
Min & 1Q & Median & 3Q & Max \\
-0.37304 & -0.09476 & 0.02507 & 0.08279 & 0.46707 \\
\hline
\multicolumn{5}{||m{10cm}||}{\small Residual standard error: 0.1401 on 135 degrees of freedom. Multiple R-squared:  0.7873, Adjusted R-squared:  0.7857, F-statistic: 499.8 on 1 and 135 DF,  p-value: $ < 2.2e-16 $} \\
\hline
\hline
\end{tabular}}
\end{center}
\end{table}

\begin{table}[ht]
\begin{center}
\caption{Fit of lineal model between sky brightness (SQM with filter V) and PM10 particle concentration.\newline}
\scalebox{0.80}{\begin{tabular}{ ||m{2cm} m{1.5cm} m{1.5cm} m{1.5cm} m{1.5cm}|| }
\hline
\hline
\small Coefficients: & \small Estimate & \small Std. Error & \small t value & \small $ Pr(>|t|) $ \\
\hline
\hline
\small (Intercept) & 18.615162 & 0.023941 & 777.54 & $ < 2e-16 $ \\
PM10 & -0.017554 & 0.000777 & -22.59 & $ < 2e-16 $ \\
\hline
\hline
\multicolumn{5}{||m{10cm}||}{Residuals:} \\
Min & 1Q & Median & 3Q & Max \\
-0.34851 & -0.09898 & 0.01589 & 0.09059 & 0.46167 \\
\hline
\multicolumn{5}{||m{10cm}||}{\small Residual standard error: 0.1359 on 135 degrees of freedom. Multiple R-squared:  0.7908, Adjusted R-squared:  0.7893, F-statistic: 510.4 on 1 and 135 DF,  p-value: $ < 2.2e-16 $} \\
\hline
\hline
\end{tabular}}
\end{center}
\end{table}

\begin{table}[ht]
\begin{center}
\caption{Fit of lineal model between sky brightness (SQM with filter B) and PM10 particle concentration.\newline}
\scalebox{0.80}{\begin{tabular}{ ||m{2cm} m{1.5cm} m{1.5cm} m{1.5cm} m{1.5cm}|| }
\hline
\hline
\small Coefficients: & \small Estimate & \small Std. Error & \small t value & \small $ Pr(>|t|) $ \\
\hline
\hline
\small (Intercept) & 20.022582 & 0.036769 & 544.55 & $ < 2e-16 $ \\
PM10 & -0.022660 & 0.001193 & -18.99 & $ < 2e-16 $ \\
\hline
\hline
\multicolumn{5}{||m{10cm}||}{Residuals:} \\
Min & 1Q & Median & 3Q & Max \\
-0.54947 & -0.11564 & 0.01529 & 0.14125 & 0.47185 \\
\hline
\multicolumn{5}{||m{10cm}||}{\small Residual standard error: 0.2087 on 135 degrees of freedom. Multiple R-squared:  0.7276, Adjusted R-squared:  0.7256, F-statistic: 360.6 on 1 and 135 DF,  p-value: $ < 2.2e-16 $} \\
\hline
\hline
\end{tabular}}
\end{center}
\end{table}

\begin{table}[ht]
\begin{center}
\caption{Fit of multivariate lineal model between sky brightness (SQM), PM10 particle concentration and time (prior to lockdown).\newline}
\scalebox{0.80}{\begin{tabular}{ ||m{2cm} m{1.5cm} m{1.5cm} m{1.5cm} m{1.5cm}|| }
\hline
\hline
\small Coefficients: & \small Estimate & \small Std. Error & \small t value & \small $ Pr(>|t|) $ \\
\hline
\hline
\small (Intercept) & 17.792510 & 0.144242 & 123.352 & $ < 2e-16 $ \\
PM10 & -0.009709 & 0.001400 & -6.936 & $ 2.10e-09 $ \\
Time & 1.792745 & 0.224364 & 7.990 & $ 2.76e-11 $ \\
\hline
\hline
\multicolumn{5}{||m{10cm}||}{Residuals:} \\
Min & 1Q & Median & 3Q & Max \\
-0.27534 & -0.06539 & 0.00570 & 0.07555 & 0.33540 \\
\hline
\multicolumn{5}{||m{10cm}||}{\small Residual standard error: 0.1161 on 66 degrees of freedom. Multiple R-squared:  0.8851, Adjusted R-squared:  0.8816, F-statistic: 254.2 on 2 and 66 DF,  p-value: $ < 2.2e-16 $} \\
\hline
\hline
\end{tabular}}
\end{center}
\end{table}

\begin{table}[ht]
\begin{center}
\caption{Fit of multivariate lineal model between sky brightness (SQM), PM10 particle concentration and time (during lockdown).\newline}
\scalebox{0.80}{\begin{tabular}{ ||m{2cm} m{1.5cm} m{1.5cm} m{1.5cm} m{1.5cm}|| }
\hline
\hline
\small Coefficients: & \small Estimate & \small Std. Error & \small t value & \small $ Pr(>|t|) $ \\
\hline
\hline
\small (Intercept) & 18.286660 & 0.088734 & 206.083 & $ < 2e-16 $ \\
PM10 & -0.009353 & 0.001365 & -6.853 & $ 3.15e-09 $ \\
Time & 0.936452 & 0.144649 & 6.474 & $ 1.46e-08 $ \\
\hline
\hline
\multicolumn{5}{||m{10cm}||}{Residuals:} \\
Min & 1Q & Median & 3Q & Max \\
-0.244612 & -0.037644 & 0.003376 & 0.069018 & 0.140332 \\
\hline
\multicolumn{5}{||m{10cm}||}{\small Residual standard error: 0.08546 on 65 degrees of freedom. Multiple R-squared: 0.7253, Adjusted R-squared: 0.7168, F-statistic: 85.79 on 2 and 65 DF,  p-value: $ < 2.2e-16 $} \\
\hline
\hline
\end{tabular}}
\end{center}
\end{table}

\begin{table}[ht]
\begin{center}
\caption{Fit of multivariate lineal model between B-band sky brightness (SQMB), PM10 particle concentration and time (prior to lockdown).\newline}
\scalebox{0.85}{\begin{tabular}{ ||m{2cm} m{1.5cm} m{1.5cm} m{1.5cm} m{1.5cm}|| }
\hline
\hline
\small Coefficients: & \small Estimate & \small Std. Error & \small t value & \small $ Pr(>|t|) $ \\
\hline
\hline
\small (Intercept) & 18.194792 & 0.158405 & 114.86 & $ < 2e-16 $ \\
PM10 & -0.008363 & 0.001537 & -5.44 & $ 8.36e-07 $ \\
Time & 2.762678 & 0.246395 & 11.21 & $ < 2e-16 $ \\
\hline
\hline
\multicolumn{5}{||m{10cm}||}{Residuals:} \\
Min & 1Q & Median & 3Q & Max \\
-0.28917 & -0.08599 & 0.00845 & 0.07619 & 0.37138 \\
\hline
\multicolumn{5}{||m{10cm}||}{\small Residual standard error: 0.1275 on 66 degrees of freedom. Multiple R-squared: 0.9067, Adjusted R-squared: 0.9039, F-statistic: 320.9 on 2 and 66 DF,  p-value: $ < 2.2e-16 $} \\
\hline
\hline
\end{tabular}}
\end{center}
\end{table}

\begin{table}[ht]
\begin{center}
\caption{Fit of multivariate lineal model between B-band sky brightness (SQMB), PM10 particle concentration and time (during lockdown).\newline}
\scalebox{0.85}{\begin{tabular}{ ||m{2cm} m{1.5cm} m{1.5cm} m{1.5cm} m{1.5cm}|| }
\hline
\hline
\small Coefficients: & \small Estimate & \small Std. Error & \small t value & \small $ Pr(>|t|) $ \\
\hline
\hline
\small (Intercept) & 18.865195 & 0.076974 & 245.084 & $ < 2e-16 $ \\
PM10 & -0.003348 & 0.001184 & -2.828 & 0.00623 \\
Time & 1.711520 & 0.125479 & 13.640 & $ < 2e-16 $ \\
\hline
\hline
\multicolumn{5}{||m{10cm}||}{Residuals:} \\
Min & 1Q & Median & 3Q & Max \\
-0.15142 & -0.06566 & 0.02261 & 0.05481 & 0.11129 \\
\hline
\multicolumn{5}{||m{10cm}||}{\small Residual standard error: 0.07413 on 65 degrees of freedom. Multiple R-squared: 0.8226, Adjusted R-squared: 0.8171, F-statistic: 150.7 on 2 and 65 DF, p-value: $ < 2.2e-16 $} \\
\hline
\hline
\end{tabular}}
\end{center}
\end{table}
  
\end{document}